\documentclass[preprint2]{aastex}
\usepackage{hyperref}

\title{Detailed X-Ray Line Properties of \thetatwo\ in Quiescence}
\shorttitle{X-ray Line Emissions of \thetatwo}
\author{Arik W. Mitschang\altaffilmark{1}, Norbert S. Schulz\altaffilmark{2}, David P. Huenemoerder\altaffilmark{2}, Joy S. Nichols\altaffilmark{1}, Paola Testa\altaffilmark{1}}
\shortauthors{Mitschang et. al.}
\altaffiltext{1}{Smithsonian Astrophysical Observatory (SAO), Cambridge, MA}
\altaffiltext{2}{MIT Kavli Institute for Astrophysics and Space Research, Cambridge, MA}

\begin{document}

\newcommand{\thetatwo}{$\theta^{2}$ Ori A}
\newcommand{\thetaonec}{$\theta^{1}$ Ori C}
\newcommand{\tetaa}{$\theta^{2}$ Ori A}
\newcommand{\tetc}{$\theta^{1}$ Ori C}
\newcommand{\tausco}{$\tau$ Sco}
\newcommand{\Msun}{M$_{\odot}$}
\newcommand{\chandra}{\textit{Chandra}}
\newcommand{\nefir}{\ion{Ne}{9}}
\newcommand{\neH}{\ion{Ne}{10}}
\newcommand{\mgfir}{\ion{Mg}{11}}
\newcommand{\mgH}{\ion{Mg}{12}}
\newcommand{\sifir}{\ion{Si}{13}}
\newcommand{\siH}{\ion{Si}{14}}
\newcommand{\ofir}{\ion{O}{7}}
\newcommand{\oH}{\ion{O}{8}}
\newcommand{\myemail}{amitschang@head.cfa.harvard.edu}
\newcommand{\kms}{km s$^{-1}$}
\newcommand{\ergsec}{ergs s$^{-1}$}
\newcommand{\R}{$\mathcal{R}$}
\newcommand{\G}{$\mathcal{G}$}
\newcommand{\RRs}{$R/R_\star$}
\begin{abstract}
\label{sec-1}

We investigate X-ray emission properties of the peculiar X-ray source
\thetatwo\ in the Orion trapezium region using more than 500 ksec of
HETGS spectral data in the quiescent state. The amount of exposure
provides tight constraints on several important diagnostics involving
O, Ne, Mg, and Si line flux ratios from He-like ion triplets,
resonance line ratios of the H- and He-like lines and line
widths. Accounting for the influence of the strong UV radiation field
of the O9.7V star we can now place the He-like line origin well within
two stellar radii of the O-star's surface. The lines are resolved with
average line widths of $341\pm$38 km s$^{-1}$ confirming a line origin
relatively close to the stellar surface. In the framework of standard
wind models this implies a rather weak, low opacity wind restricting
wind shocks to temperatures not much larger than 2$\times10^6$ K.  The
emission measure distribution of the X-ray spectrum, as reported
previously, includes very high temperature components which are not
easily explained in this framework.  The X-ray properties are also not
consistent with coronal emissions from an unseen low-mass companion
nor with typical signatures from colliding wind interactions. The
properties are more consistent with X-ray signatures observed in the
massive Trapezium star \tetc\ which has recently been successfully
modeled with a magnetically confined wind model.
\end{abstract}
\section{Introduction}
\label{sec-2}
\tetaa\ is a triple star system at the heart of the Orion Nebula
Cluster (ONC), Its massive primary has been identified as a 5th
magnitude O9.5 V star \citep{abt91} with a mass of 25 \Msun\ 
\citep{preibisch1999}, making it the second most massive star in the
ONC next to the 45 \Msun\ O5.5 V star of \tetc. A more recent
photometric study provides an optical identification of O9 V and a
total system mass of 39$\pm14$ \Msun\ \citep{simon2006}.  The studies
of \citet{abt91} and \citet{preibisch1999} show that this system
includes two close intermediate mass companions at 174 AU and 0.47 AU
separation with mass estimates between 7 and 9 \Msun\ for each.

\tetaa\ has been extensively monitored in X-rays with The \chandra\ 
X-ray Observatory and has shown its fair share of odd
behavior. Observations in 2000 found that the X-ray source exhibited
unusual and dramatic variability with a 50$\%$ flux drop in less than
12 hours accompanied by multiple small flares with only a few hours
durations \citep{feigelson2002}. Such behavior in an early type
stellar system is surprising since this can not be explained by the
standard wind shock models for X-rays in early type stars
\citep{lucy1982, owocki1989}, nor by the magnetically confined wind
models (MCWMs; \citet{babel1997}). While the MCWM can produce hard
X-ray emission like observed in \tetc\ \citep{schulz2000, schulz2003,
gagne2005} and \tausco\ \citep{cohen2001}., it does not explain the
observed variability in \thetatwo.  At the time, the suggestion was
made that such emission could be the result of magnetic reconnection
events. To add to this excitement, a specifically powerful X-ray flare
from \thetatwo, seen with the \chandra\ High Energy Transmission
Grating Spectrometer (HETGS), surprised observers in 2004
\citep{schulz2006} and produced a total power output exceeding
10$^{37}$ \ergsec. Considering the orbital phase of the close
spectroscopic companion, the low He-like forbidden/intercombination
line ratios, and the fact that all lines remained unresolved led to
the argument that these events are triggered by magnetic interactions
with the close companion. A sub-pixel re-analysis of a similar flare
event which appeared during observations in the \chandra\ Orion
Ultradeep Project (COUP) \citep{stelzer2005, schulz2006} in 2003,
however, seem to indicate that these events may originate from the
companion instead (M. Gagne, priv. comm.). An unseen T Tauri companion
appears unlikely due to the observed peculiar line properties.

In contrast to that observed in the elevated states, the quiescent
spectrum of \thetatwo\ exhibits temperatures above 25 MK and has line
ratios which suggest that the X-ray emitting plasma is close enough to
the stellar surface of the massive star to argue for some form of
magnetic confinement \citep{schulz2006}.  The argument is strengthened
by the fact that the line widths, quite in contrast to the narrow line
widths observed during the outbursts, seem broadened to the order of
300 \kms. These properties are very reminiscent of the MCWM results
obtained in \tetc\ \citep{gagne2005}, where, through detailed
simulations, it was demonstrated that the bulk of the emitting plasma
is close to the photosphere, or within $\sim$2 R$_{\star}$, and line
widths are $\leq$400 \kms. However, in spite of these apparent
differences, the properties of the quiescent state remained fairly
unconstrained with respect to precise line ratios and widths. \tetaa's
X-ray luminosity is about an order of magnitude lower than than that
observed during outburst and the study remained statistically limited.

The \chandra\ Data Archive (CDA)\footnote{\href{http://cxc.harvard.edu/cda/}{http://cxc.harvard.edu/cda/}} now
contains an additional $\sim$ 300 ks on \tetaa\ between 2004 and 2008
and in this paper we present a full analysis of the quiescent spectrum
allowing us to derive much better constrained line properties. The
results are also used to test the hypothesis that the X-ray emission
from \tetaa\ is consistent with predictions from the MCWM. The paper
is structured as follows; in Section \ref{:sec:obs:} we discuss the
observations and analysis methods, in Section \ref{:sec:discuss:} we
discuss the results of our emission line measurements, and finally we
summarize our findings in Section \ref{:sec:conclusion:}
\section{Observations and Analysis \label{:sec:obs:}}
\label{sec-3}
\begin{figure*}[tbp] \centering\ 
\includegraphics[scale=0.7,angle=270]{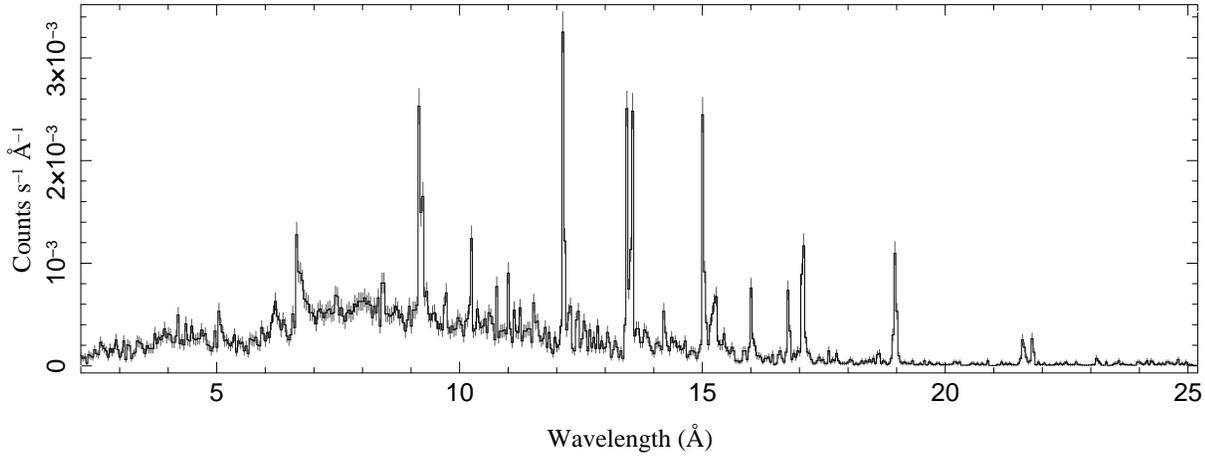}
\caption{Counts spectrum from the total combined data on
\thetatwo\ (MEG+HEG). \label{:fig:fullspectrum:}}
\end{figure*}

We have retrieved \chandra\ HETGS data in the vicinity of the ONC, which were
originally observed as a part of the HETG Orion Legacy Project
\citep{holp}, from the CDA. There are now seventeen separate \chandra\ 
observations which include \thetatwo\ within an off-axis angle suitable
for extraction. See Table \ref{:tbl:obsids:} for a list of the
included observations and selected properties. Noting that this study
is focused on the quiescent state spectrum and that ObsID 4474 was
not included in any analysis in the current study due to the
substantially elevated count rate during its entire exposure, we have
accumulated 520 ks of exposure time on \thetatwo\ in the quiescent
state. Figure \ref{:fig:fullspectrum:} shows the total combined counts
spectrum using the 520 ks on \thetatwo.
\begin{deluxetable}{lrccccr}
\tablecaption{Observation Log \label{:tbl:obsids:}}
\tablewidth{0pt}
\tablehead{
\colhead{Sequence} &
\colhead{ObsID} &
\colhead{Start Date} &
\colhead{Start Time} &
\colhead{Exposure} & 
\colhead{Offset\tablenotemark{a}} &
\colhead{Phase Range\tablenotemark{b}}\\
\colhead{Number} &
\colhead{} &
\colhead{(UT)} &
\colhead{(UT)} &
\colhead{($ks$)} &
\colhead{arcmin} &
\colhead{}
 }
\startdata
200001 & 3                    & 1999-10-31 & 05:58:56 & 49.6 & 2.42 & 0.76-0.79\\
200002 & 4                    & 1999-11-24 & 05:39:24 & 30.9 & 2.28 & 0.92-0.99\\
200175 & 2567                 & 2001-12-28 & 12:25:56 & 46.4 & 1.98 & 0.99-1.01\\
200176 & 2568                 & 2002-02-19 & 20:29:42 & 46.3 & 2.10 & 0.53-0.55\\
200242 & 4473                 & 2004-11-03 & 01:48:04 & 49.1 & 1.26 & 1.00-1.03\\
200243 & 4474\tablenotemark{c}& 2004-11-23 & 07:48:38 & 50.8 & 1.39 & 0.96-0.99\\
200420 & 7407                 & 2006-12-03 & 19:07:48 & 24.6 & 1.64 & 0.27-0.29\\
200423 & 7410                 & 2006-12-06 & 12:11:37 & 13.1 & 3.02 & 0.40-0.41\\
200421 & 7408                 & 2006-12-19 & 14:17:30 & 25.0 & 2.08 & 0.02-0.04\\
200422 & 7409                 & 2006-12-23 & 00:47:40 & 27.1 & 2.30 & 0.19-0.21\\
200424 & 7411                 & 2007-07-27 & 20:41:22 & 24.6 & 3.94 & 0.53-0.54\\
200425 & 7412                 & 2007-07-28 & 06:16:09 & 25.2 & 4.39 & 0.55-0.56\\
200462 & 8568                 & 2007-08-06 & 06:54:08 & 36.1 & 2.53 & 0.98-1.00\\
200462 & 8589                 & 2007-08-08 & 21:30:35 & 50.7 & 2.53 & 0.10-0.13\\
200478 & 8897                 & 2007-11-15 & 10:03:16 & 23.7 & 3.37 & 0.80-0.81\\
200477 & 8896                 & 2007-11-30 & 21:58:33 & 22.7 & 2.34 & 0.54-0.55\\
200476 & 8895                 & 2007-12-07 & 03:14:07 & 25.0 & 1.74 & 0.84-0.85\\
\enddata 
\tablenotetext{a}{\thetatwo\ zeroth order position offset from nominal pointing}
\tablenotetext{b}{Assuming a 20.974 day period and periastron passage at
HJD$_{\circ}$=2440581.27 \citep{abt91}}
\tablenotetext{c}{4474 is included here only for reference, no analysis herein utilized
it due to the extremely elevated count rate during its entirety}
\end{deluxetable}


As noted, none of these observations were targeted at \thetatwo;
indeed no \chandra\ gratings observations have ever targeted
\thetatwo. However using the suite of advanced extraction tools
provided by the \chandra\ Transmissions Grating Catalog and Archive
(TGCat; \citet{tgcat, tgcatproceedings})\footnote{\href{http://tgcat.mit.edu}{http://tgcat.mit.edu}}, extraction of the
dispersed counts of off-axis X-ray source positions proved to be
trivial.

Grating spectra were extracted and responses computed using TGCat
software to locate the optimal centroid position of \thetatwo\ and
apply proper calibration for each observation.  In a crowded field
such as the Orion Trapezium, careful attention must be made during
analysis to contamination from other zeroth order counts lying close
to or on top of dispersion counts and dispersion arms crossing one
another at critical points. To this end, we reviewed order sorting
images (ACIS CCD event energy vs. gratings order $\times$ wavelength,
or specifically FITS-file columns $\mathrm{TG\_MLAM}$
vs. $\mathrm{ENERGY}$) for each observation and identified potential
contamination. In this view, the source traces two hyperbolas centered
on $m\lambda=0$ (e.g. see \chandra\ 
POG\footnote{\href{http://cxc.harvard.edu/proposer/POG/}{http://cxc.harvard.edu/proposer/POG/}} Fig 8.13); traces from
confusing sources show as offset hyperbolas (dispersed) or vertical
lines (zeroth order).  We found no significant source of contamination
in the regions used for line fitting; See Table \ref{:tbl:lines:} for
details on the locations of these regions. Similarly when fitting the
continuum we used a set of wavelength ranges containing few lines, the
``line free regions'', in which we found little contamination. See
Section \ref{:sec:continuum:} for a more detailed discussion on the
continuum modeling and Section \ref{:sec:linefit:} on line
fitting. Line width analysis is treated separately in Section
\ref{:sec:linewidths:}.

All fitting of data was done using the Interactive Spectral
Interpretation System
(ISIS; \citet{isisref})\footnote{\href{http://space.mit.edu/CXC/isis}{http://space.mit.edu/CXC/isis}}, along with
the Astrophysical Plasma Emission Database
(APED; \citet{apedref})\footnote{\href{http://cxc.harvard.edu/atomdb/sources\_aped.html}{http://cxc.harvard.edu/atomdb/sources\_aped.html}}
for line emissivities and continuum modeling.
\subsection{Continuum \label{:sec:continuum:}}
\label{sec-3_1}

The continuum emission of \thetatwo\ was modeled by fitting a single
temperature APED model to the combined MEG+HEG counts for all
observations to improve statistics. In order to fit only the continuum
emission, we selected a set of narrow bands, considered free of
significant line emission, specifically 2.00-2.95\AA{}, 4.4-4.6\AA{},
5.3-6.0\AA{}, 7.5-7.8\AA{}, 12.5-12.7\AA{} and 19.1-20 \AA{} (e.g. see
\citet{testa2007}). We assumed a hydrogen column density (N$_H$) of
$2\times10^{21}$ cm$^{-2}$. Potential contamination resulting from
cross-dispersion or zeroth order confusion was mitigated in these
regions by simply ignoring the affected region of an individual order
during the computation of the fit. The resulting continuum model was
then used when fitting lines.
\subsection{Line Fluxes \& Ratios \label{:sec:linefit:}}
\label{sec-3_2}
\begin{figure*}
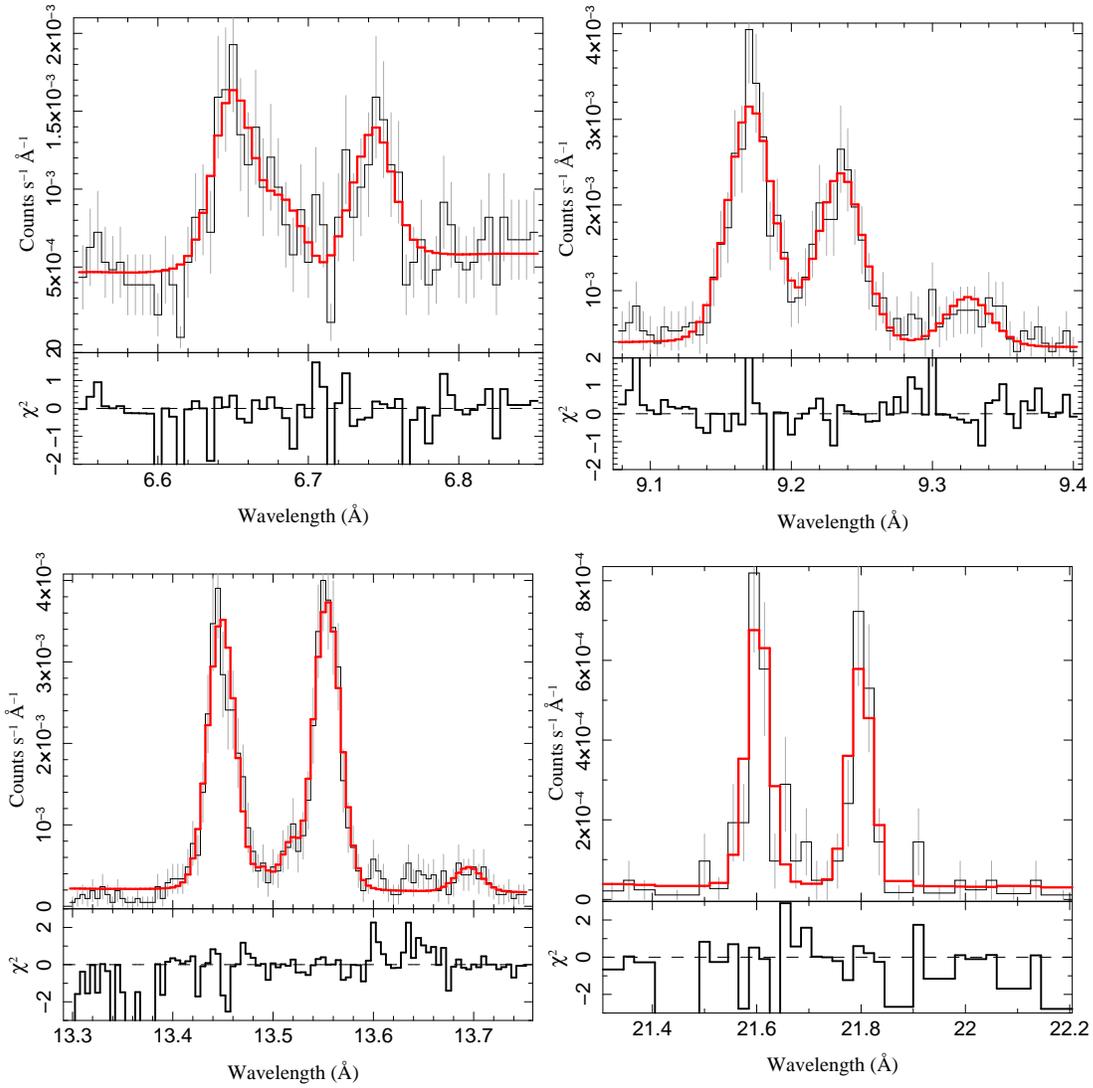
 \centering\ 
\includegraphics[scale=0.40,angle=270]{si13-GR-fit.ps}
\includegraphics[scale=0.40,angle=270]{mg11-GR-fit.ps} \\
\includegraphics[scale=0.40,angle=270]{ne9-GR-fit.ps}
\includegraphics[scale=0.40,angle=270]{o7-GR-fit.ps} 
\caption{Fit to the fir triplet \G- and \R-ratios for, clockwise from
top left, \sifir, \mgfir, \ofir, \nefir\ showing the predicted line
profile in red, data and errors in black and gray respectively. The
line centroid positions for each $fir$ component are given here for
clarity [$r$,$i$,$f$]: \sifir\ [6.65,6.68,6.74], \mgfir\ 
[9.17,9.24,933], \ofir\ [21.60,21.80,22.11] and \nefir\ 
[13.45,13.55,13.69].  \label{:fig:fits:}}
\end{figure*}

\begin{figure*}
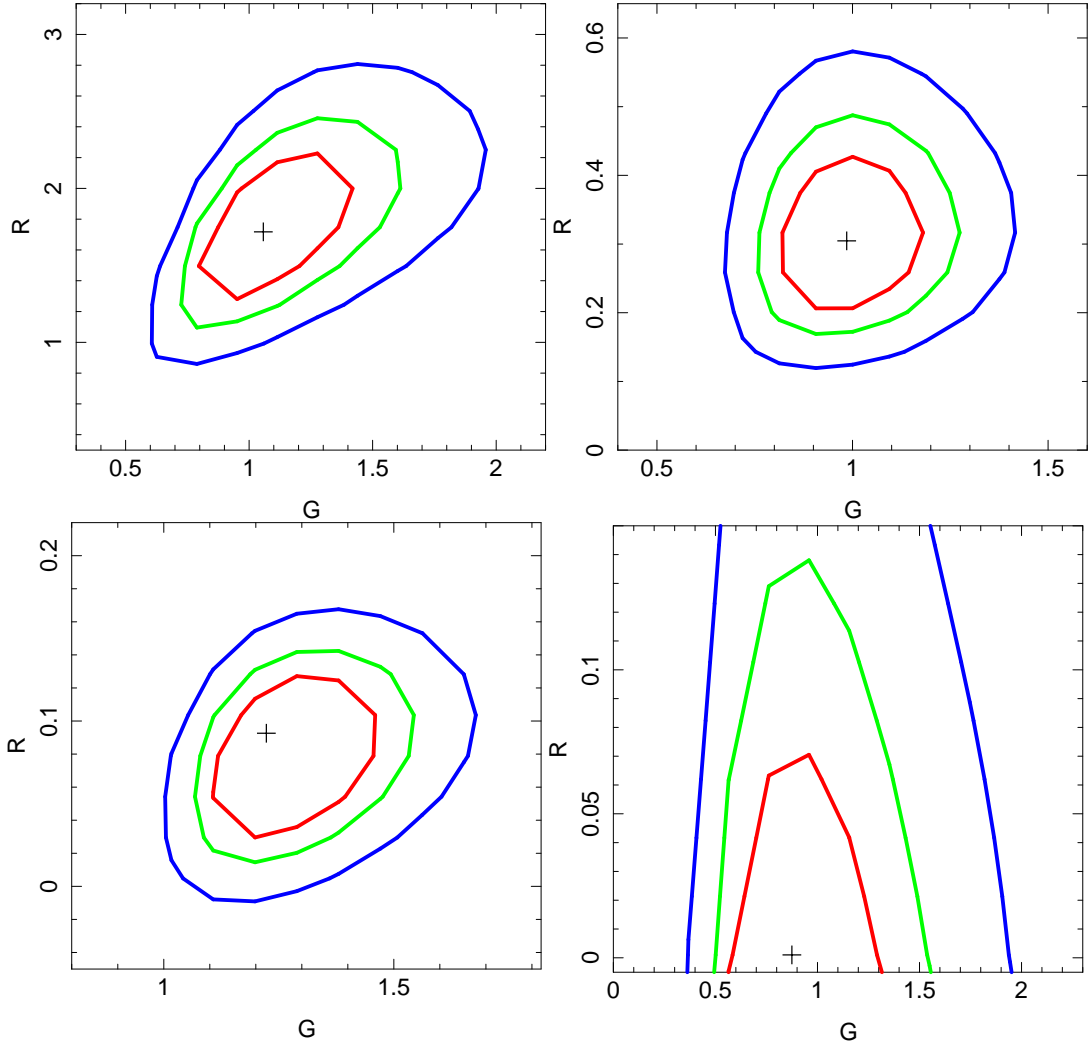

\centering\ 
\includegraphics[scale=0.40,angle=270]{si13-GR-conf.ps}
\includegraphics[scale=0.40,angle=270]{mg11-GR-conf.ps} \\
\includegraphics[scale=0.40,angle=270]{ne9-GR-conf.ps}
\includegraphics[scale=0.40,angle=270]{o7-GR-conf.ps}
\caption{
Confidence contours for measured \G- and \R-ratios for, clockwise
from top left, \sifir, \mgfir, \ofir, \nefir. The red inner contours
show 1$\sigma$, green middle contours show 2$\sigma$ and outer blue
contours show 3$\sigma$ confidences. \label{:fig:conf:}}
\end{figure*}
The \textit{fir} (forbidden, intercombination, and resonance) line
ratios given by \R$=f/i$ and \G$=(f+i)/r$ have been shown to be probes
of both density (\R) and temperature (\G) \citep{gabjord69} in X-ray
emitting plasmas, and in the presence of a strong UV radiation field,
such as is typical in O stars like \thetatwo, \citet{walcas01}
demonstrated that the \R\ value rather acts as a proxy for the radial
distance of X-ray emission from the stellar surface.  Specifically for
the \R-ratio, it is also important to make the comparison between the
observed ratio and that of the low density limit. \citet{blumenthal1972}
showed that

\begin{equation}\label{eq:ro} 
\mathcal{R} = \frac{\mathcal{R}_\circ}{1 + \frac{\phi}{\phi_c} + \frac{n}{n_c}}
\end{equation} 

where $\phi/\phi_c$ is a measure of the photo-excitation, $n/n_c$ is a
measure of the density and \R$_o$, $\phi_c$, and $n_c$ depend only on
atomic parameters and temperature. It is easily seen from
Eq. \ref{eq:ro} that, ignoring photo-excitation, \R$_o$=\R\ when
$n/n_c\ll1$ and thus represents the low density limit. We have
computed \R$_o$ using emissivities in APED and temperatures derived from
the \G\ ratio given in Table \ref{:tbl:hhetemp:} for each $fir$ triplet
and list them in Table \ref{:tbl:lines:}. In order to test the MCWM
predictions we derive the radial distance (\RRs) using these $fir$
ratios, a surface temperature of 30,000 K for the 09.7V star and
photo-excitation and decay rates from \citet{blumenthal1972}. Figure
\ref{:fig:rvsrstar:} shows the dependence curves with measured values
and 90\% confidence intervals over-plotted.

In cases where there were significant contributions from other lines
or line groups, as in the case of \nefir\ $fir$ triplet where
\ion{Fe}{19} and \ion{Fe}{21} converge and blend, those lines were
included in the model.  A special case is \neH\ which is unresolvably
blended with FeXVII. In this case we assumed the Fe component
contributed flux equaling 13\% of the flux of a prominent FeXVII line
at 15.01\AA{} (e.g. see \citet{walborn2009}). Additionally, the Mg
Ly-series converges at the centroid position of the \sifir\
\textit{f}-line where we assumed, based on the theoretical relative
line strengths, the observed flux was overestimated by 10\% of the
measured flux of the isolated H-Like \ion{Mg}{12} Ly$\alpha$ line.

When fitting the He-like $fir$ triplet lines, the relative separation
of the lines was fixed and the positions of the resonance lines were
constrained by their rest positions. Where available, we fit using
both MEG and HEG counts, where MEG counts were rebinned onto the HEG
grid whose intrinsic channel size is half that of the MEG. Fits were
performed by applying Gaussian functions for each contributing
line. Several contaminating lines known to be in the vicinity were
included as well.  The \G- and \R-ratios were computed directly during
the fitting procedure and the $fir$ fluxes were treated
co-dependently.  The instrumental profile was included as calibration
data while the excess width was included as a gaussian turbulent
broadening term ($v_\mathrm{turb}$).  In Figures \ref{:fig:fits:} and
\ref{:fig:conf:} the triplet regions are shown with residuals,
over-plotted models, and computed confidence contours.
\subsection{Line Widths \label{:sec:linewidths:}}
\label{sec-3_3}
Due to degradation of \chandra\ image quality at off-axis angles, the
HETG resolving power likewise decreases. Though the PSF is well
defined across the ACIS detector, this degradation becomes a problem
for gratings because, owing to the complexity of modeling, responses are
only calibrated for zeroth order positions at the instrument nominal
pointing.

This effect can be critical in line width measurements which may
include a significant instrumental broadening signature. Our flux
measurements are unaffected by the broadening, and we have utilized as
much available data as possible to improve statistics. Four of our
observations are at off-axis angles greater than the others, in
particular \thetatwo\ is greater than 3$^\prime$ off-axis in obsids
7410, 7411, 7412, and 8897. We have chosen to ignore counts in these
obsids during computation of line width parameters.  There are two
exceptions. \siH\ and \mgH\ where statistics are too poor in the absence
of extra counts to obtain reasonable measurements. In these cases we
provide upper limits on the line widths.

The average offset of our data is 2$^\prime$.1 which is around the
location where degradation becomes noticeable. Based on analysis of
ACIS zeroth order Line Response Functions (LRFs) at large axial
offsets (e.g. see \chandra\ POG), we estimate that our reported line
widths are on the order of up to $\sim$5\% broader than that of
identical on-axis profiles.
\begin{deluxetable}{lrccccl}
\tablecaption{Line Measurements \label{:tbl:lines:}}
\tablewidth{0pt}
\tablecolumns{7}
\tablehead{
\colhead{ION} & 
\colhead{$\lambda$\tablenotemark{a}} &
\colhead{flux\tablenotemark{b}} &
\multicolumn{3}{c}{Line Ratios\tablenotemark{c}} &
\colhead{$v_{\mathrm{turb}}$} \\
\colhead{} &
\colhead{(\AA)} &
\colhead{($10^{-6} \mathrm{phot s^{-1} cm^{-2}}$)} &
\colhead{$\mathcal{G}$} &
\colhead{$\mathcal{R}$} &
\colhead{\R$_o$\tablenotemark{d}} &
\colhead{($\mathrm{km s^{-1}}$)}
}
\startdata
\multicolumn{7}{c}{He-Like Lines} \\
\hline
\sifir & 6.650  &  1.3$\pm$0.3    & 1.1$\pm$0.3 & 1.7$\pm$0.4  & 2.4 & 491$\pm$120 \\
\mgfir & 9.171  &  4.8$\pm$0.6    & 1.0$\pm$0.1 & 0.3$\pm$0.1  & 3.1 & 432$\pm$53  \\
\nefir & 13.448 & 28.9$\pm$3.3    & 1.2$\pm$0.1 & 0.1$\pm$0.04 & 2.8 & 228$\pm$34  \\ 
\ofir  & 21.602 & 147.3$\pm$40.9  & 0.9$\pm$0.3 &      $<$0.09 & 4.1 & 274$\pm$83  \\

\cutinhead{\hspace{0.5in}H-Like Ly$\alpha$ Lines\hspace{0.4in}$\frac{\mathrm{H\ Ly}\alpha}{\mathrm{He\ Ly}\alpha}$}

\siH   & 6.187  &  0.4$\pm$0.2    & \multicolumn{3}{c}{0.3$\pm$0.2} & $<$686      \\ 
\mgH   & 8.423  &  0.8$\pm$0.3    & \multicolumn{3}{c}{0.2$\pm$0.1} & $<$518      \\
\neH   & 12.133 &  20.0$\pm$2.9   & \multicolumn{3}{c}{0.7$\pm$0.1} & 315$\pm$43  \\
\oH    & 18.971 & 164.7$\pm$22.3  & \multicolumn{3}{c}{1.1$\pm$0.4} & 327$\pm$53  \\

\enddata
\tablenotetext{a}{Measured position of resonance line for He-like triplet line groups}
\tablenotetext{b}{flux is that of the resonance line only for He-like triplet line groups}
\tablenotetext{c}{for H-Like Ly$\alpha$ lines this is the ratio of the
H-Like Ly$\alpha$ flux to He-Like resonance line flux of the
corresponding ion}
\tablenotetext{d}{Computed from APED emissivities according to
Eq. \ref{eq:ro} at \G-ratio derived temperatures (see Table
\ref{:tbl:hhetemp:})}
\end{deluxetable}

\section{Discussion \label{:sec:discuss:}}
\label{sec-4}
The exposure obtained from the \chandra\ archive of \thetatwo\
represents the deepest combined high resolution spectroscopic dataset
on this young massive O-star to date. The long exposure provides high
statistics in critical emission lines, allowing to diagnose its X-ray
stellar wind properties beyond the $3\sigma$ level. In a previous
study \citet{schulz2006} provided some preliminary results for the
quiescent state for less than half of the current exposure. This
limited measurements of critical line fluxes and widths to
uncertainties larger than 50$\%$. Our new analysis greatly reduces
these uncertainties to the order of 20$\%$. For example, while the
previous analysis could only speculate about possible line broadening
of the order of 300 km s$^{-1}$, we now clearly resolve the lines to
values between 228$\pm$34 km s$^{-1}$ for \nefir\ and 491$\pm$120 km
s$^{-1}$ for \sifir, with an average of all lines of 341$\pm$38 km
s$^{-1}$. Likewise critical line ratios such as the \R-ratios are
significantly improved, specifically for the cases of \mgfir\ with
0.3$\pm$0.09 and \sifir\ with 1.7$\pm$0.4; for the case of \ofir\,
since its $f$-line was not detected, we now also have an upper limit.

\begin{figure}[h] \centering\ 
\plotone{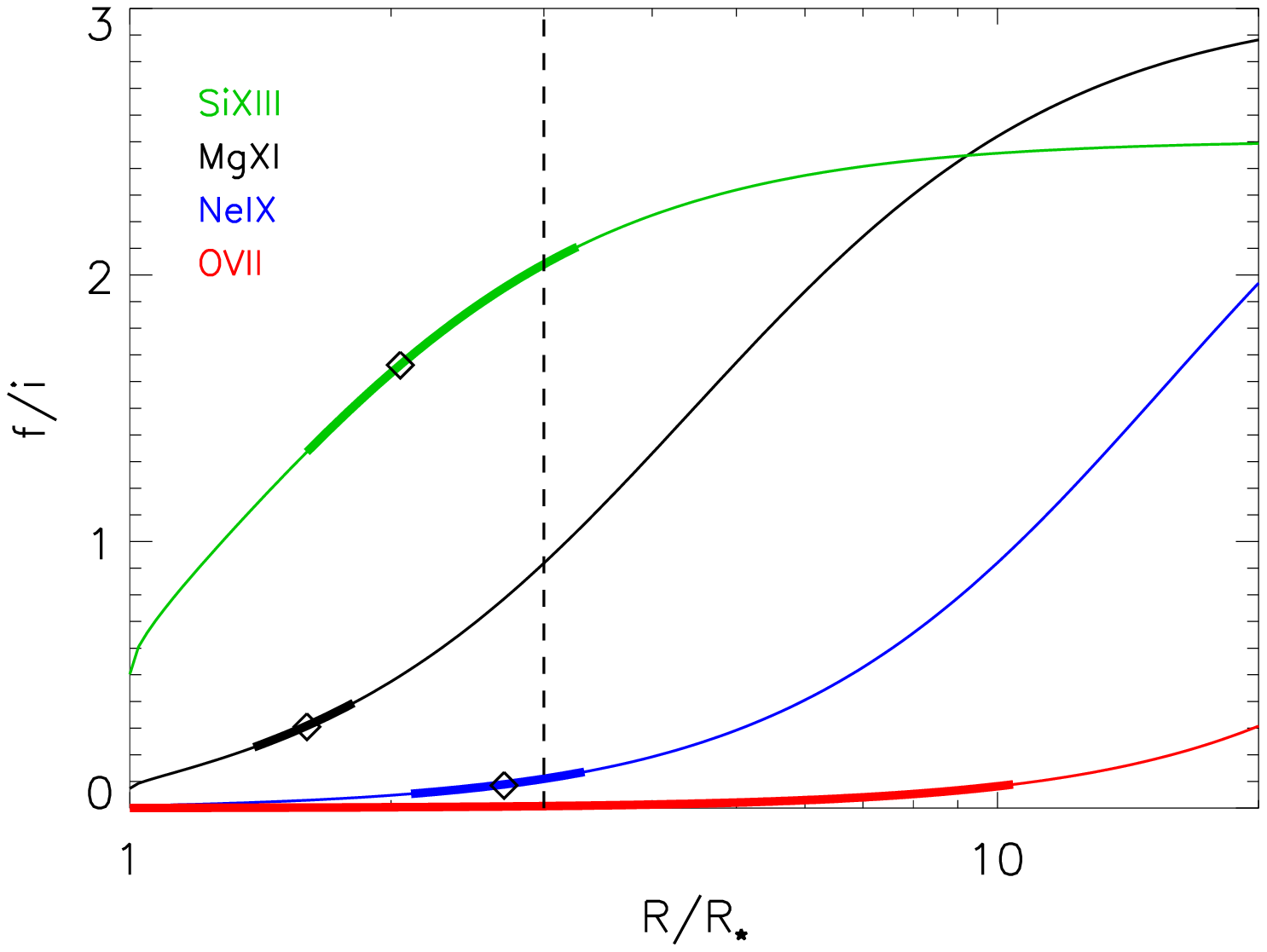} 
\caption{Dependence of \R\ ratios on the distance to the stellar
surface of the emission from \sifir, \mgfir, \nefir, \ofir. The
diamonds show the best fit and highlighted lines show 90\%
confidence limits projected on to dependence curves computed for a
stellar surface temperature of 30,000 K. The vertical dashed line at
2R$_\star$ represents the approximate theoretical limit for generation
of X-rays under the MCWM.\label{:fig:rvsrstar:}}
\end{figure}

The measured \R-ratios are significantly less than \R$_o$ (Table
\ref{:tbl:lines:}). In early type stars this is due to the substantial
UV radiation field provided by blackbody radiation of the hot surface
temperature \citep{kahn2001, gabjord69}, which for the O9.7V star in
\thetatwo\ is about 30,000 K. In this case the \R\ ratio maps the
distance of emission from the stellar surface and we utilize this
relation to show that the X-ray line emission from \thetatwo\ is
indeed located close to the O-star's surface.  \citet{schulz2006}
estimated that the emissions could be within several stellar radii,
Table \ref{:tbl:lines:} and Figure \ref{:fig:rvsrstar:} show emission
origins within two stellar radii (dotted line in Figure
\ref{:fig:rvsrstar:}) for \sifir, \mgfir, and \nefir\ with their
90$\%$ uncertainties.
\begin{deluxetable}{lllll}
\tablecaption{Derived Temperatures\label{:tbl:hhetemp:}}
\tablewidth{0pt}
\tablecolumns{3}
\tablehead{
\colhead{ION} &
\colhead{Log T(H/He)} &
\colhead{Log T(G)}
}
\startdata
\ofir  & 6.41 (6.36, 6.44) & 6.3 (6.0, 6.5) \\
\nefir & 6.62 (6.60, 6.64) & 6.1 (6.0, 6.3) \\
\mgfir & 6.70 (6.65, 6.72) & 6.5 (6.3, 6.6) \\
\sifir & 6.96 (6.87, 7.02) & 6.4 (6.0, 6.8) \\
\enddata
\end{deluxetable}

Another important result of our analysis is that the measured line
centroid positions shown in Table \ref{:tbl:lines:} are, with quite
high accuracy, at the expected ion rest wavelengths indicating that
there are no line shifts within the \chandra\ sensitivity. This is an
important result because any shift would indicate fast outward moving
sources in a high density wind. The line profiles appear symmetric,
supporting a low density wind assumption even though at such low
broadening, profile deviations are almost impossible to trace even at
our data quality.

In the case of $\zeta$ Ori, \citet{walcas01} find that lines are
resolved with comparatively low Doppler velocities of around 900 km
s$^{-1}$, \R-ratios that are characteristic of several stellar radii,
an extremely small \sifir\ \R-ratio, and symmetric and unshifted
lines. Except for the extremely small \sifir\ \R-ratio, our results
seem very similar, if not more extreme with respect to line widths and
\R-ratios. Our line widths indicate an even lower shock jump velocity
than in the case of $\zeta$ Ori making the formation of the observed
ionization states even more difficult.  At 350 km s$^{-1}$ shock
temperatures are expected to not exceed 2$\times10^6$ K. This
discrepancy is supported by the emissivity distribution of the
spectrum which includes X-ray temperatures greater than 25 MK
\citep{schulz2006}.  These results are difficult to reconcile within
the standard wind model.  In this respect we conclude that a picture
of a low density wind with shocks produced near its onset is not
particularly convincing.

There are not many scenarios left which could explain our findings.
We can rule out significant contributions of unseen low-mass pre-main
sequence companions by the level of the line broadening. Standard
coronal emission would show unresolved lines or moderate broadening
due to orbital motion \citep{brickhouse2001, huenemoerder2006};
neither is the case here. Colliding winds are ruled out simply by the
fact this would require an unseen massive companion with a much
earlier type than the O9.7, which would be impossible to hide.

We find, however, quite strong similarities to the most massive star
in the Orion Trapezium \thetaonec\ \citep{schulz2003, gagne2005}.  In
the magnetically confined wind scenario, field lines of the magnetic
dipole act to channel emitted material from either pole toward the
magnetic equator. Simulations by \citet{gagne2005} demonstrate that
these two components meet at the magneto-equator and wind plasma with
high tangential velocities reaching up to 1000 km s$^{-1}$ collides
generating strong shocks and elevate gas temperatures to tens of
millions of degrees, thus producing the observed hard X-ray
emission. \citet{gagne2005} further demonstrate that the conditions
for X-ray production are quite specific; the post shock in-falling
material is rather cool, and the outflowing material's density is too
low to produce sufficient X-rays. This places a relatively tight
constraint on the location of the hard X-ray emission around
R$\leq$2R$_\star$.

Another result of the simulations by \citet{gagne2005} states that
the post shock-heated material is moving slowly, thus generating
observed line profiles much narrower than expected for non-magnetic
shock-heated X-ray production in O stars \citep{lucy1982, walcas01}. In
order to quantify the expected broadening, \citet{gagne2005} recreated
emission measure and line profiles from the simulations and found that
the turbulent broadening is expected to be on the order of 
250 \kms, with little to no blueshift in the line centroid position, which
is very close to what we observe in \thetatwo.
\section{Conclusion \label{:sec:conclusion:}}
\label{sec-5}

We have analyzed high resolution X-ray spectra from \chandra\ on the
young massive O star \thetatwo, totaling over 500 ks in the quiescent
state, and computed line widths and $fir$ line ratios for a series of
prominent emission lines appearing in its spectrum. The resulting
measurements show relatively narrow lines at an average width of
341$\pm$38 \kms\ and \R-ratio derived X-ray emitting origin within 2
stellar radii. Comparing these results to the simulation results of
\citet{gagne2005} for \tetc, we argue that the X-ray production
mechanism in \thetatwo\ is most likely via magnetic confinement of its
stellar wind outflows.

We have explored other possibilities, including standard O-star wind
models and close companions, for the the generation of X-rays in
\thetatwo\ but find that none of these are ideal for explaining the
observed spectral properties. Observed line widths are too low, while
shock temperatures too high to satisfy model predictions in most of
these cases.

Finally, we note that this is a comparative analysis, and sets up a
case for a more rigorous analysis specifically aimed at
magneto-hydrodynamical modeling (MHD) using the MCWM similar to that
under-taken for \tetc.
\acknowledgments\ 
\label{sec-6}

This research has made use of data obtained from the Chandra Data
Archive and software provided by the Chandra X-ray Center (CXC) in the
application package CIAO. This research also made use of the Chandra
Transmission Grating Catalog and archive
\href{http://tgcat.mit.edu}{http://tgcat.mit.edu}. \chandra\ is operated by the Smithsonian
Astrophysical Observatory under NASA contract NAS 8-03060.  This work
was supported by NASA through the Smithsonian Astrophysical
Observatory (SAO) contracts NAS 8-03060 and SV3-73016 for the Chandra
X-Ray Center and Science Instruments.

Facilities: \facility{Chandra}
\bibliographystyle{apj}
\bibliography{theta2}
\end{document}